# Temperature dependence of the interface spin Seebeck effect

Farhan Nur Kholid[a,b], Dominik Hamara[a], Marc Terschanski[c], Fabian Mertens[c], Davide Bossini[d], Mirko Cinchetti[c], Lauren McKenzie-Sell[a,e], James Patchett[a], Dorothée Petit[a], Russell Cowburn[a], Jason Robinson[e], Joseph Barker[f], Chiara Ciccarelli[a*]

[a]*Cavendish Laboratory, University of Cambridge, Cambridge, CB3 0HE, United Kingdom*
[b]*Physics Department, Lancaster University, Lancaster, LA1 4YB, United Kingdom*
[c]*Experimentelle Physik VI, TU Dortmund, Otto-Hahn-Strasse 4, D-44227 Dortmund, Germany*
[d]*Department of Physics and Center for Applied Photonics, University of Konstanz, Universitätstraße 10, 78457 Konstanz, Germany*
[e]*Department of Materials Science and Metallurgy, 27 Charles Babbage Rd, Cambridge CB3 0FS, United Kingdom*
[f]*School of Physics and Astronomy, University of Leeds, Leeds, LS2 9JT, United Kingdom*

*cc538@cam.ac.uk

**We performed temperature-dependent optical pump – THz emission measurements in $Y_3Fe_5O_{12}$ (YIG)|Pt from 5 K to room temperature in the presence of an externally applied magnetic field. We study the temperature dependence of the spin Seebeck effect and observe a continuous increase as temperature is decreased, opposite to what is observed in electrical measurements where the spin Seebeck effect is suppressed as 0K is approached. By quantitatively analysing the different contributions we isolate the temperature dependence of the spin-mixing conductance and observe features that are correlated to the bands of magnon spectrum in YIG.**

The longitudinal spin Seebeck effect (LSSE)[1] describes the transfer of a spin current from a magnetic insulator driven by a temperature gradient. An adjacent heavy metal (HM) layer with large spin orbit coupling is typically used to convert the spin current into an electrical signal via the inverse spin Hall effect (ISHE).[2,3] The LSSE has been measured in a variety of different materials such as ferromagnets[1,4,5], anti-ferromagnets[6,7] and paramagnets.[8] Magnetic insulators (MI) such as $Y_3Fe_5O_{12}$ (Yttrium Iron Garnet – YIG) are particularly interesting for studies on the LSSE since the absence of electron charge transport allows the roles of magnons and phonons to be identified in the spin transfer.[1,3,9,10] Temperature, thickness and magnetic field dependence studies have contributed to a phenomenological picture of magnon-driven spin current.[11–15] A temperature gradient across the magnetic insulator thickness leads to the diffusion of thermal magnons that accumulate at the interface with the HM.[16,17] The temperature dependence of the magnon propagation length $\lambda_\mathrm{m}$ results in a characteristic peak in the SSE signal at low temperature when the thickness of the MI is comparable to $\lambda_\mathrm{m}$.[12] Low frequency magnons play a dominant role due to their large population and longer thermalisation lengths. Their contribution can be suppressed by large magnetic fields which raise the energies of the magnon spectrum.[14,15]

This picture of a bulk-like transport induced by a temperature gradient picks up the essential features of the LSSE. However, several experimental results raise questions on the details of how the spin current is transferred across at the MI|HM interface.[12] This contribution has been challenging to isolate in electrical measurements of the LSSE and its temperature dependence is not known.

Recently, ultra-fast experimental techniques using femtosecond lasers have enabled the study of the LSSE and the underlying physical mechanisms of spin current generation at picosecond and shorter timescales.[18,19] In these experiments a laser pulse rapidly heats the free electrons in the HM, quickly thermalising to an effective temperature, $T_\mathrm{e}$. The temperature of the magnons in the insulator, $T_\mathrm{m}$, is increased primarily by the spin current which propagates across the interface from the hotter metal. This thermalisation processes is proportional to $T_\mathrm{e} - T_\mathrm{m}$ and its timescale is ultimately determined by the electron-magnon scattering time.[18] In this ultra-short time window after the laser excitation a thermal gradient is not yet established in the bulk of the MI and the spin current generation originates only at the interface between MI and HM.[19]

In this study, we measured the LSSE in YIG|Pt on the picosecond timescale in the low temperature range from 5 K to room temperature. We observed a different temperature dependence of the LSSE compared to DC electrical studies carried out in the same temperature range[12,14,15]. Our sample is a 100 nm thick commercial YIG film grown by liquid phase epitaxy on a (111)-oriented Gd$_3$Ga$_5$O$_{12}$ (GGG) substrate. We cleaned the surface using piranha etching and then sputtered a 5 nm thick layer of Pt on top. Fig. 1 shows the two different orientations of our experiments. We pump the sample from either the GGG side or the Pt side with 50 fs laser pulses with a central wavelength of 800 nm. Any spin transfer across the YIG|Pt interface triggered by the pump pulse is converted into an electric current via the inverse spin-Hall effect in the Pt layer. This produces a broad-band electric-dipole emission $E_\mathrm{THz}(\omega)$ with a bandwidth directly related to the Fourier transform of the spin current $j_\mathrm{s}(\omega)$ as[20]

$$E_\mathrm{THz}(\omega) = \frac{Z_0}{n_\mathrm{YIG}(\omega)+n_0(\omega)+\int_0^d Z_0 \sigma_\mathrm{Pt}(\omega)\mathrm{d}z} \frac{\lambda_\mathrm{s} \Theta_\mathrm{SH} e j_\mathrm{s}(\omega)}{\hbar} \quad (1)$$

where $Z_0$ is the free space impedance in Ohms, $\hbar$ is Planck's constant, $e$ is the charge of an electron, $\lambda_\mathrm{s}$, $\sigma_\mathrm{Pt}$, $d$ and $\Theta_\mathrm{SH}$ are respectively the spin diffusion length in nm, the electrical conductivity in Ohms$^{-1}$ cm$^{-1}$, the thickness in nm, and the spin-Hall angle of the Pt layer. $n_\mathrm{YIG}(\omega)$ and $n_0(\omega)$ represent the refractive indices of YIG and air. We detect the emitted radiation by electro-optic sampling with a 1-mm thick ZnTe crystal. The detected signal $S(\omega)$ is the convolution of Eq. (1) with the detector response function, which is bandwidth limited to 0.2-2.5 THz range. We apply an external magnetic field ($\mu_0 H = \pm 0.5$ T) along the [100] direction (Fig. 1) during the measurements to saturate the YIG magnetisation. We extract an odd-in-magnetic field $S_- = [S(+H) - S(-H)]/2$ and an even-in-magnetic field $S_+ = [S(+H) + S(-H)]/2$ contribution to the overall emission. $S_+$ is polarised in the [100]-[010] plane (Fig. 2a and 2b). Its dependence on the pump polarisation (Fig. 2a) connects its origin to optical rectification. Both bulk GGG and YIG are centrosymmetric.[21,22] However, their lattice mismatch induces elastic deformations in YIG close to the interface that gradually changes its lattice parameters, breaking inversion symmetry and yielding a non zero value for the second order electro-optic constant $\chi^{(2)}$, as also confirmed by the measurement of optical second harmonic generation.[23] From this point forward, we focus on the $S_-$ contribution that is due to the LSSE. Unlike $S_+$, $S_-$ does not show any dependence on the pump polarisation and is always polarised along the [010] axis, perpendicular with respect to the interface normal and the YIG magnetisation (Fig. 2b). The reversal of the interface normal vector with respect to the pump pulse propagation direction results in a polarity switching of the emitted THz radiation (Fig. 2c). Both observations are consistent with the symmetry of the ISHE for a spin current travelling across the interface with spin polarisation along the [100] direction.[2] As a function of the external magnetic field, the peak amplitude of $S_-$ follows the hysteresis curve of the YIG magnetisation (Fig. 2d), also in agreement with previous electrical and optical measurements of the LSSE.[18,24]

Fig. 3a shows the temperature dependence of $S_-$. The continuous line represents a fitting with the function $(T_C - T)^\alpha$, where $T_C = 550$ K is the Curie temperature and $\alpha = 2.9 \pm 0.1$. This trend is similar to the temperature dependence measured above room temperature with both low-frequency electrical[11] and ultra-fast optical methods[18], but is remarkably different from the low temperature behaviour of the LSSE measured in adiabatic conditions, where the signal diminishes towards 0 K.[12,14] In our experiment we detect the spin current generated in a time interval up to a few picoseconds after the laser absorption in Pt. This interval is orders of magnitude shorter than the time needed to establish a thermal gradient in bulk YIG (1-100 nanoseconds).[25,26] Thus, we are probing the electron-magnon interactions localised at the interface. The temperature difference emerges between the free electron system on the HM side, $T_e$, and the spin system on the MI side, $T_m$. This temperature difference is significantly larger than that generated in a DC electrical experiment. Thermalisation between these two systems occurs via direct electron-magnon interaction which is the origin of the spin transfer across the interface.[18,19] The interfacial spin transport parameters are summarised by the spin-mixing conductance $g^{\uparrow\downarrow}$ and the resulting spin current can be written as[17,27]

$$j_s = \frac{\gamma \hbar k_B g^{\uparrow\downarrow}}{2\pi M_s V}(T_e - T_m) \quad (2)$$

where $\gamma$ is the gyromagnetic ratio, $k_B$ is the Boltzmann constant, $M_s$ is the saturation magnetisation of YIG and $V$ is the unit cell volume. In the case of a femtosecond laser excitation, $T_e - T_m$ is set by the energy deposited in the HM layer, in other words by the absorbed laser fluence. The Pt layer has a strong optical absorption ($\sim 10^7 \text{cm}^{-1}$)[28] that is enhanced by the Etalon effect[29], while the absorption in GGG|YIG (10 cm$^{-1}$) is essentially negligible.[30,31] We estimate that for an absorbed fluence of 0.15 mJ/cm² $\Delta T_{e,\max} \sim 200$ K at 10 K[18], taking the electron-phonon coupling constant $g_{e-ph} = 10^{18}$ W·m$^{-3}$K$^{-1}$ and the electron heat capacity $C_e/T_e = 750$ J·m$^{-3}$K$^{-2}$.[32]

To understand the origin of the temperature dependence of the picosecond LSSE in Fig. 3a, we consider all parameters that contribute to its magnitude, as expressed by Eq. (1) and (2). We first normalise $S_-$ with the YIG magnetisation $1/M_s$ obtained from SQUID measurement (Fig. 3b), which shows that $M_s$ is not accountable for the large change in the THz emission. To experimentally verify how $\Delta T_e$ is influenced by the ambient temperature, we perform pump-probe transient reflectivity measurement on glass|Pt bilayers from 10 K to 300 K. The transient change in reflectivity $\Delta R/R(t)$ is proportional to $\Delta T_e(t)$.[33,34] As seen in Fig. 3c, The peak magnitude of $\Delta R/R$ is weakly affected by the ambient temperature. Similarly, the time evolution of $\Delta R/R$ is also independent of the ambient temperature with the thermalisation time between electrons and phonons $\tau_{e-ph} = 260 \pm 10$ fs, obtained by an exponential fit $\propto \exp\left(-\frac{t}{\tau_{e-ph}}\right)$. This measurement indicates that we have approximately the same temperature difference, $T_e - T_m$, at all ambient temperatures used in our experiment.

Apart from $g^{\uparrow\downarrow}$, which quantifies the quality of the interface in conducting spins, the other parameters ($\lambda_s$, $\sigma_{Pt}$ and $\Theta_{SH}$) are intrinsic to the Pt layer. To exclude the contribution of these transport parameters or any other contribution from the set-up, we compare our LSSE results to a metallic THz spintronic emitter[20,35] CoFeB (3 nm)|Pt (5 nm). In this case, the pump beam hits the sample from the CoFeB side and is largely absorbed by the ferromagnet, inducing a strong superdiffusive spin current.[36,37] Therefore, far from $T_C = 1100$ K[38], CoFeB behaves as a temperature-independent spin current source, transported to the Pt layer by high mobility majority spin carriers. Eq. (1) also applies to this metallic bilayer as it relies on the spin-to-charge conversion in Pt to generate THz emission. In agreement with a previous report[39], the amplitude of the THz pulse decreases with decreasing temperature and reaches a plateau at 50 K (Fig. 3d). This behaviour, which is associated with the intrinsic components of the spin Hall effect in Pt[39], significantly differs to what is observed in our YIG|Pt sample, allowing us to exclude the influence of the Pt layer in our measured temperature dependence of the LSSE. We conclude therefore that our measurement probes the temperature dependence of the spin mixing conductance.

The laser-excited free electrons in Pt are not spin polarised initially. The stochastic local exchange field fluctuations induced by single electron scattering events off the interface with the MI are therefore averaged to zero at timescales longer than the interaction time (~ 4fs for YIG|Pt[18]). Higher order interactions between the scattering electrons and the MI can lead to a net magnetic torque on the MI and therefore to spin accumulation, as described in Ref. 18. An additional contribution associated with the real part of the spin-mixing conductance $g^r$ is given by inelastic spin-flip scattering processes that result in the excitation of a magnon on the MI side. This contribution depends on the density of states of magnons as well as the electronic density of states at $T_e$. Using Eqs (1) and (2) we estimate the range of the spin mixing conductance at 10 K as $g^{\uparrow\downarrow} = (1.8 - 8.4) \times 10^{18}$ m$^{-2}$, in agreement with that found in [14]. Our parameters are $Z_0 = 377$ Ω, $n_{YIG} = 5$[18], $n_o = 1$, $\sigma_{Pt}(10\ K) = 0.03\ \mu\Omega^{-1}\text{cm}^{-1}$ [39], $\lambda_s(10\ K) = 2 - 4$ nm [40,41], $\Theta_{SH} = 0.01 - 0.0223$ [40], $M_s(10\ K)$ = 172 kA/m, $V = a^3$, $a$ =1.24 nm [42], $T_e - T_m \approx 200$ K. Note that $n_{YIG}$ and $\sigma_{Pt}$ can be considered frequency-independent within our detection bandwidth[28,43]. We postulate that the kink in the temperature dependence of the LSSE signal around 80 K (Fig. 3b) may be related to the higher energy magnon bands in YIG becoming populated by the highly energetic electrons in Pt. At an ambient temperature of 100 K the first high-frequency bands appear at ~25 meV [44-46], which coincides with the average energy of the optically heated electrons in Pt. The progressive filling of these bands at higher ambient temperature affects the spin pumped across the interface and determines the temperature dependence of the LSSE.

In conclusion, we characterise the low temperature behaviour of the picosecond spin Seebeck effect in YIG|Pt by optical pump-THz emission measurents and show that it is substantially different from that reported in low-frequency electrical measurements. We observe a sustained increase of the signal with decreasing temperature, which is a continuation of the previous femtosecond SSE experiment measured from room temperature to above $T_c = 550$ K. This behaviour cannot be attributed to a variation of the temperature gradient at the interface or of the spin and charge transport characteristics in Pt, and is instead to be associated with the spin-mixing conductance, providing direct access to its temperature dependence.

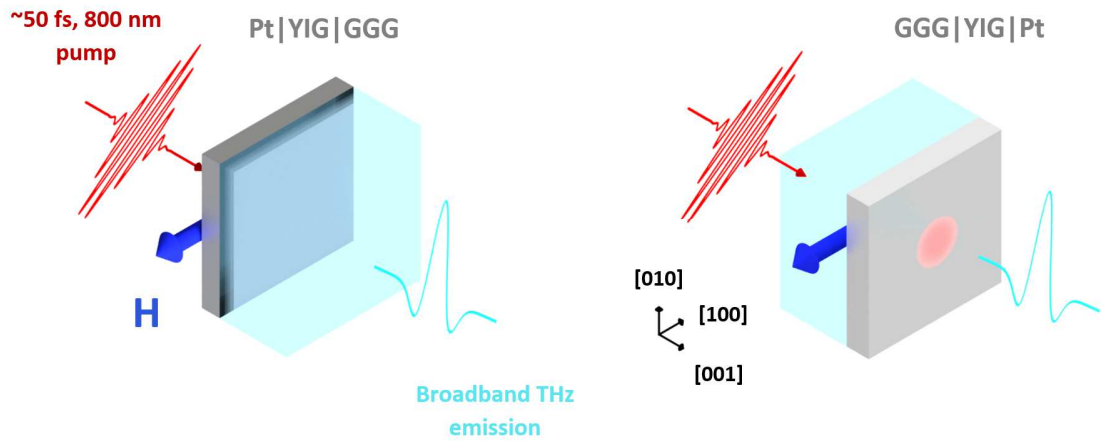

**FIG. 1.** Schematic illustration of the experiment performed with the femtosecond laser pulses incident on the Pt side (left) and the GGG substrate side (right).

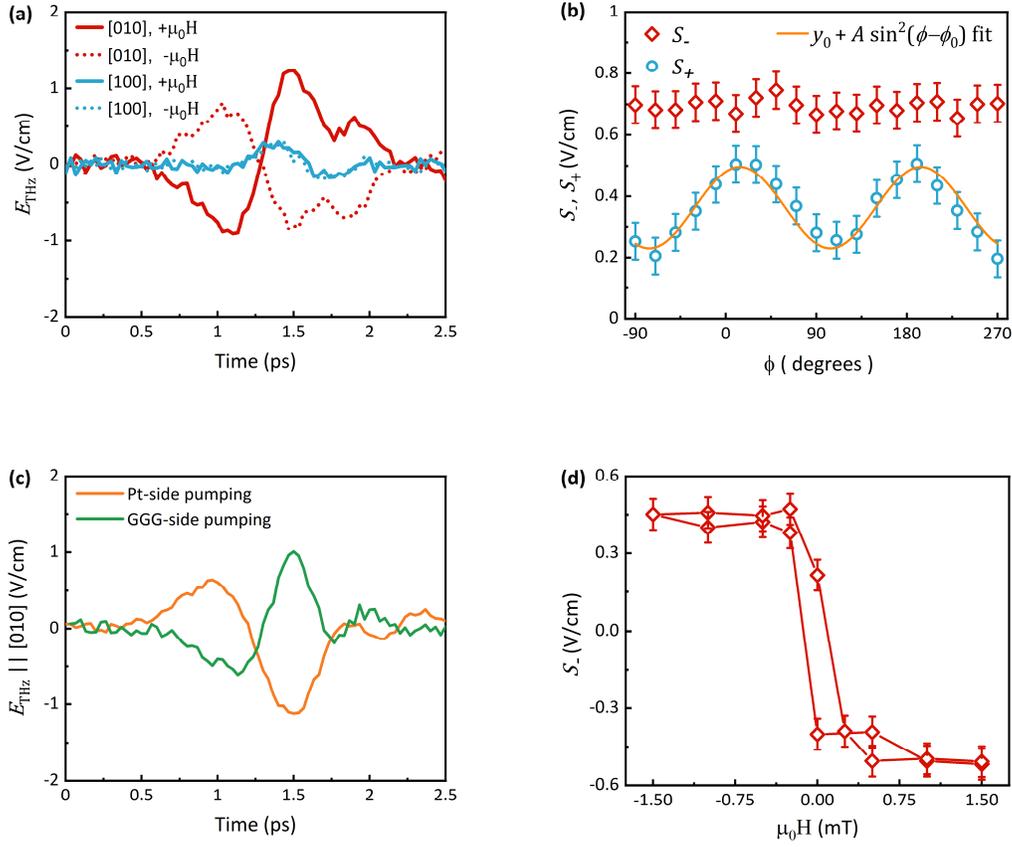

**FIG. 2.** (a) Time-domain THz emission resolved along the [100]- and [010]-axis, measured at 10 K. Odd-in-field signal $S_-$ only appears along the [010]-axis component whereas the signal along the [100]-axis is even-in-field $S_+$ ($\mu_0 H = \pm 0.5$ T). (b) $S_+$ (blue circle) and $S_-$ (red diamond) dependence on the linear pump polarisation where the angle $\phi$ is relative to the [010]-axis. These measurements were carried out for signals along [010]-axis at room temperature. The orange line is a fit using $y_0 + A \sin^2(\phi - \phi_0)$, where $y_0$ is a constant offset, $A$ is the magnitude of the optical rectification signal, $\phi_0$ is an angle offset. This angular dependence agrees with the 2$^{nd}$ harmonic generation measurement in GGG|YIG[23]. An offset of -0.3 V/cm is applied to $S_+$ for clarity. (c) $E_{\text{THz}}$ polarised along [010]-axis in time-domain for Pt-side and GGG-side pumping. (d) Hysteresis curve of $S_-$ measured at room temperature.

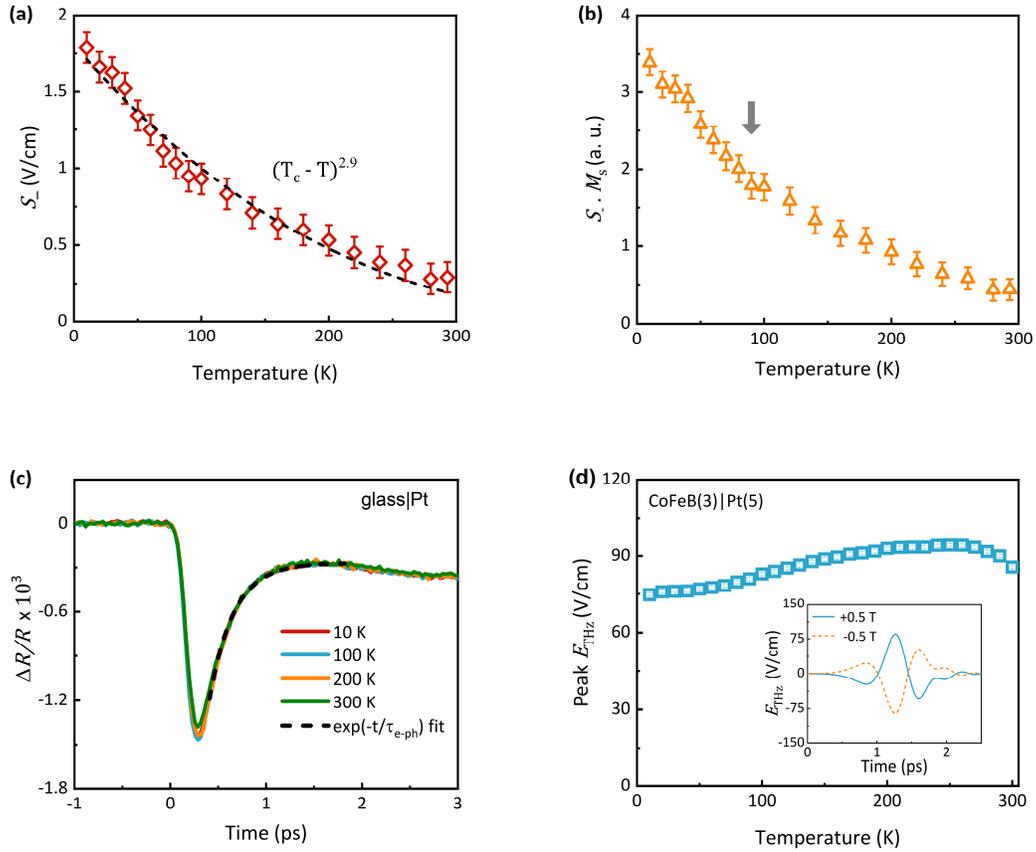

**FIG. 3.** (a) Temperature dependence of $S_-$ The dashed line is a fit with a function $A(T_C - T)^\alpha$, $A = 1.5 \pm 1.3$, $\alpha = 2.9 \pm 0.1$, and $T_C = 550$ K. (b) Temperature dependence of the normalised $S_-$ with the YIG magnetisation $M_s$. The grey arrow indicates the temperature at which the slope $dS_-/dT$ changes (c) Time-resolved transient reflectivity ($\Delta R/R$) of glass|Pt measured in a temperature range of 10 – 300 K at a fixed pump fluence of 0.4 mJ/cm². The dashed line is an exponential fit $\propto \exp\left(-\dfrac{t}{\tau_{e-ph}}\right)$ (d) Peak THz emission from CoFeB(3 nm)|Pt(5 nm) as a function of ambient temperature where the pump pulse hits from the CoFeB side. The error bar is comparable with the symbol size. The inset shows the time-domain data for opposite field polarities $\pm 0.5$ T.

**Acknowledgement**

F.N.K. acknowledges support from the Cambridge Trust and the Jardine Foundation. D.H acknowledges support from the Winton Programme for the Physics of Sustainability. J.P.P. acknowledges support from the EPSRC. M.C., D.B., F.M. and M.T. acknowledge support from the Deutsche Forschungsgemeinschaft through the International Collaborative Research Center TRR160 (Projects B8 and B9). C.C. and J. B. acknowledge support from the Royal Society and the Winton Programme for the Physics of Sustainability.